\documentclass[aip,apl,reprint]{revtex4-1}

\usepackage{graphicx}
\usepackage{dcolumn}
\usepackage{bm}
\usepackage{color}
\usepackage{verbatim} 

\begin{document}

\title{Andreev nanoprobe of half-metallic CrO$_{2}$ films using superconducting cuprate tips}

\author{C. S.~Turel}
\affiliation{Department of Physics, University of Toronto, 60 St. George Street, Toronto ON M5S1A7 
Canada}

\author{I. J.~Guilaran}
\affiliation{Department of Physics, Union University, Jackson, TN 38305 USA}

\author{P.~Xiong}
\affiliation{Department of Physics, Florida State University, Tallahassee FL USA}

\author{J. Y.T.~Wei}
\affiliation{Department of Physics, University of Toronto, 60 St. George Street, Toronto ON M5S1A7 
Canada}
\affiliation{Canadian Institute for Advanced Research, Toronto, ON, M5G1Z8 Canada}

\begin{abstract}

Superconducting tips of YBa$_{2} $Cu$_{3}$O$_{7-x}$ were used to perform point-contact Andreev reflection spectroscopy on half-metallic CrO$_{2}$ thin films. At 4.2K, strong suppression of the $d$-wave Andreev reflection characteristics was observed, consistent with the high spin polarization of CrO$_{2}$. Our technique was validated by comparison with data taken on non-magnetic Au films, and with data taken by superconducting Pb tips.  The point contacts were estimated to be $\lesssim$ 10nm in size, attesting to their ballistic and microscopic nature. Our results demonstrate the feasibility of using superconducting cuprate tips as spin-sensitive nanoprobes of ferromagnets.

\end{abstract}



\maketitle

Andreev reflection (AR) is the process by which an electron incident from a normal metal (N) is converted into a Cooper pair in a superconductor (S) \cite{BTK}.  In the case of $s$-wave pairing, AR is sensitive to the electron spin polarization in the metal counterelectrode, as a direct consequence of spin conservation \cite{deJong:PRL}.  For a normal metal, where there is an equal density of spin-up versus spin-down states at the Fermi level $E_{f}$, a spin-up electron can be retroreflected as a spin-down hole to form a spin-singlet pair thus doubling the conductance across the NS junction.  For a half metal, where the electrons at $E_{f}$ are  100\% spin-polarized, such retroreflection is inhibited thus suppressing the enhancement of junction conductance.  This inherent spin sensitivity of AR has been exploited to determine the spin polarization in a variety of itinerant ferromagnets, by measuring the conductance spectra of both point contact and fixed planar junctions \cite{Soulen:science,Upadhyay:3247}.

In the case of superconductors with $d$-wave pairing, AR can also involve quasiparticle interference and result in the formation of zero-energy bound states at the NS interface.  Basically, because of the order-parameter sign change across $d$-wave line nodes, consecutively Andreev-reflected quasiparticles can constructively interfere to produce a zero-bias peak (ZBP) in the conductance spectrum on non-principal axis junctions \cite{Hu:94PRL1526,KT:95PRL351}.  Since AR is inherently spin-dependent, this ZBP is expected to be suppressed for a ferromagnetic counterelectrode depending on the extent of its spin polarization \cite{spinpol_BTK,Zutic:PRB,Zutic:PRB2}.  Such ZBP suppression effect has been previously studied in fixed planar junctions for high-$T_c$ cuprate superconductors \cite{Vasko:844,Chen:212508} but never in point contact junctions.

In this letter we used superconducting tips of YBa$_{2} $Cu$_{3}$O$_{7-x}$ (YBCO) to perform point contact spectroscopy on ferromagnetic thin films of CrO$_{2}$, in order to study how the $d$-wave AR characteristics on YBCO are affected by the electron spin polarization of CrO$_{2}$.   While YBCO is known to have a predominantly $d$-wave pairing symmetry \cite{Tsuei_revu}, CrO$_2$ is believed to be an exemplary half-metal, with nearly perfect spin polarization \cite{CrO2_Parker}.  In order to validate our technique and interpretation, we compared conductance spectra measured on YBCO/CrO$_{2}$ junctions with spectra taken on YBCO/Au and Pb/CrO$_{2}$ junctions.  Spectra with ZBPs were observed on YBCO/Au, while spectra with zero-bias dips (ZBD) were observed on YBCO/CrO$_{2}$ and Pb/CrO$_{2}$. These observations provide direct evidence for the suppression of $d$-wave Andreev states by spin polarization in point-contact junctions.  Our point-contact radius was estimated to be $\approx$ 0.7 - 6.0 nm, demonstrating that superconducting cuprate tips can potentially be used to probe electron spin polarization by AR spectroscopy at the nanoscale.

Epitaxial thin film samples of CrO$_{2}$, $\approx$ 200 - 250 nm thick, were fabricated on (100)-oriented TiO$_{2}$ substrates using a chemical vapor deposition growth technique \cite{Ranno:JAP,Li:JAP,Gupta:APL,Ivanov:JAP,Anguelouch:JAP}.  To gauge the half metallicity of our CrO$_{2}$ film surfaces, Pb tips were used as a conventional $s$-wave superconductor for measuring the spin polarization of our films by $s$-wave AR.  Measurements were made in a $^{4}$He dipper probe between 4.2 and 8.5 K.  Differential conductance $dI/dV$ versus voltage $V$ spectra were obtained using a four-point geometry with standard ac lock-in technique. The CrO$_{2}$ films we measured had resistances much smaller than the point-contact resistances, thus ruling out any issues of spreading resistance \cite{Woods:PRB}.

\begin{figure}[t]
\centering
\includegraphics[width=3.3in]{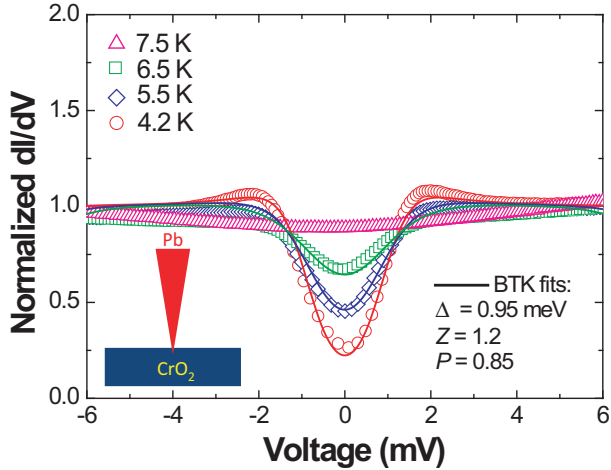}
\caption{Normalized differential conductance versus bias voltage spectrum taken on a Pb/CrO$_{2}$ point-contact junction at different temperatures.  Open symbols correspond to the spectral data, and solid lines are fits using the BTK model.}
\label{PbCrO2PCS}
\end{figure}

Figure \ref{PbCrO2PCS} shows temperature evolution of the $dI/dV$ spectra measured on a Pb/CrO$_{2}$ point-contact junction.  The spectrum at each temperature was normalized relative to the $dI/dV$ taken at energies higher than $\Delta^{Pb}$, the superconducting energy gap of Pb.  At 7.5K, above the $T_{c}$ of Pb, the $dI/dV$ spectrum shows negligible dependence on $V$.  As temperature is lowered below $T_{c}$, the subgap $dI/dV$ is progressively suppressed.  The spectral data is fitted to the modified Blonder-Tinkham-Klapwijk (BTK) model accounting for barrier strength $Z$ and spin polarization $P$ \cite{Strijkers:104510, footnote}.  The parameters used in the fit shown in Fig.\ref{PbCrO2PCS} are $\Delta^{Pb}$=0.95meV, $Z$=1.2 and $P$=0.85.  This large spin polarization is consistent with previous point-contact measurements of CrO$_{2}$ \cite{Soulen:science,Ji:PRL}, even though our $P$ value is slightly smaller, as can be explained by our relatively larger $Z$ \cite{Woods:PRB,Ji:PRL}.

Having confirmed the near-half metallicity of our CrO$_{2}$ films using $s$-wave superconducting tips, the effect of spin polarization on $d$-wave Andreev states can be determined by measuring YBCO/CrO$_{2}$ junctions.  YBCO tips were fabricated by cutting slivers, typically 2x2x5 mm$^{3}$, from a YBCO crystal monolith grown in a melt-zone furnace.  The YBCO slivers were mechanically polished into a fine tip, nominally pointed along the (110) axis. After ultrasonic cleaning in ethanol, the YBCO tips were re-annealed at 500$^{\circ}$C in flowing oxygen for 36 hours.  Before measuring YBCO/CrO$_{2}$ junctions, the YBCO tips were tested on normal-metal Au films to ensure that ZBPs due to $d$-wave Andreev interference were observed in the $dI/dV$ spectrum.  Several YBCO tips were used for the Au/YBCO point-contact junctions, whose resistance ranged from $\approx$ 10 $\Omega$ to 500 $\Omega$ at 4.2 K.    

\begin{figure}[t]
\centering
\includegraphics[width=3.3in]{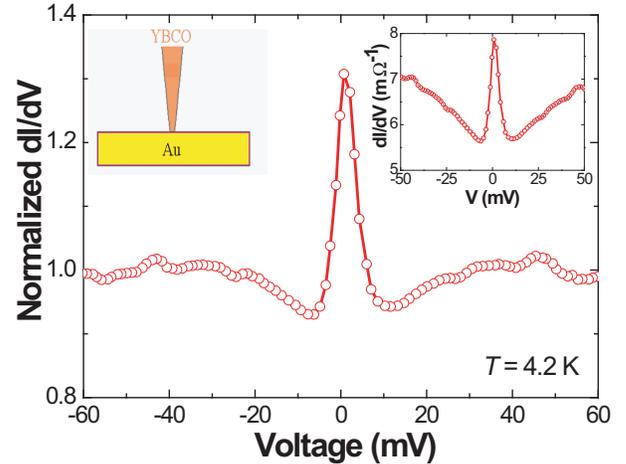}
\caption{Normalized conductance spectrum measured on a Au film using a YBCO tip at 4.2 K.  Right inset is a plot of the unnormalized spectrum showing the linear background which is characteristic of YBCO.}
\label{AuYBCO}
\end{figure}

Figure \ref{AuYBCO} shows the normalized $dI/dV$ spectrum measured on a typical Au/YBCO point-contact junction at 4.2 K, and the unnormalized data is shown in the top right inset.  The $dI/dV$ data was normalized by dividing out, using a polynomial fit, the spectral background which is often observed in YBCO junctions \cite{Geerk:329,Sun:3266}.  In the normalized spectrum, a pronounced ZBP is present along with a gap-like structure within $\sim$ $\pm$ 20mV, which is consistent with the superconducting energy-gap maximum of optimally-doped YBCO \cite{Wei:PRL}.  Such ZBP structures have been commonly observed on YBCO for non-principal axes junctions, and attributed to $d$-wave Andreev interference \cite{Wei:PRL,Deu:Rewview}.  

\begin{figure}[t]
\centering
\includegraphics[width=3.3in]{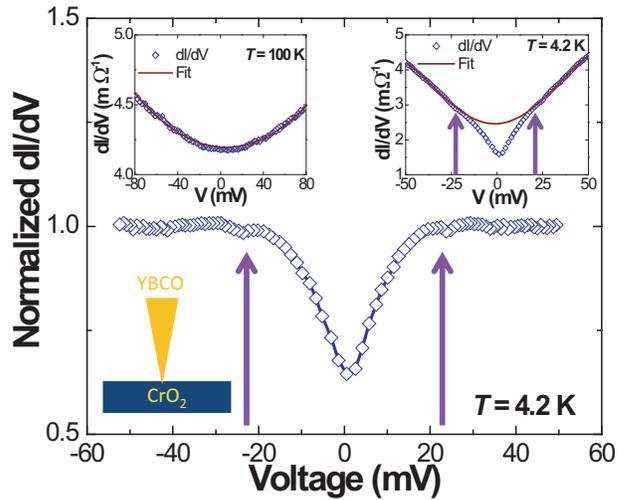}
\caption{Normalized conductance spectrum measured on a CrO$_{2}$ film using a YBCO tip at 4.2 K.  Insets show the unnormalized differential conductance spectra taken at 4.2K (right) and at 100K (left).  Open symbols represent the data while solid lines are a polynomial fit to the background.  Arrows indicate spectral kinks, whose locations are consistent with the superconducting gap maximum for YBCO.}
\label{CrO2YBCO}
\end{figure}

Figure \ref{CrO2YBCO} shows the normalized $dI/dV$ spectrum measured on a typical YBCO/CrO$_{2}$ film junction at 4.2 K, and the unnormalized data is shown in the top right inset.  The $dI/dV$ data was normalized by fitting the spectral background beyond $\pm$ 20mV to a polynomial and dividing the entire spectrum by the fit. For $|V|$ $\gtrsim$ 20mV, the normalized $dI/dV$ is relatively independent of voltage.  For $|V|$ $\lesssim$ 20mV, a ZBD is clearly observed.   Noticeable in both the normalized and unnormalized spectra are spectral kinks at $\pm$ 22mV, where the slope of $dI/dV$ shows an inflection, as indicated by the arrows in the inset.  The position of these kinks can be related to $\Delta_0$ the superconducting gap maximum of optimally-doped YBCO, signaling a crossover into the subgap regime where $dI/dV$ becomes suppressed by the spin polarization of CrO$_{2}$.

\begin{figure}[t]
\centering
\includegraphics[width=3.3in]{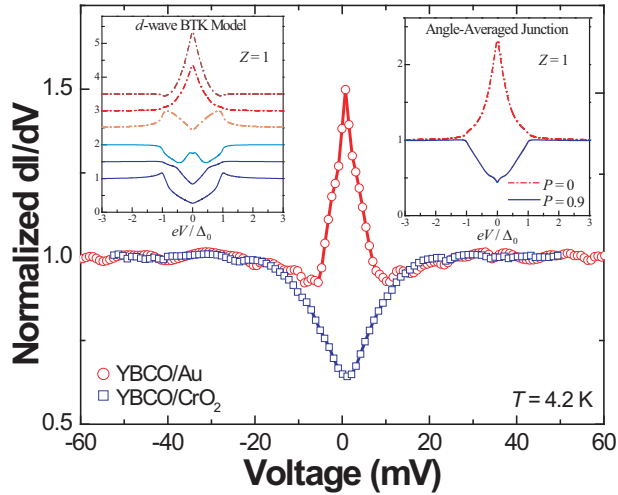}

\caption{Comparison of the normalized conductance spectra taken on YBCO/Au (circles) and YBCO/CrO$_{2}$ (squares) at 4.2 K.  Left inset shows various spectra calculated using the spin-dependent $d$-wave BTK model, for three junctions oriented normal to the $ab$-plane at $Z$=1: upper three curves are for $P$=0, with the junction normal rotated by 0, $\pi$/12 and $\pi$/4 (top to bottom) from the $d$-wave node axis; the lower three curves are for $P$=0.9 at the same three junction angles.  Right inset shows two angle-averaged spectra ($P$=0 for upper, $P$=0.9 for lower), each averaged within a Gaussian envelope of width $\pi$/6 about the $d$-wave node axis, to simulate our nominally-oriented (110) YBCO tip junctions.}
\label{AuCrO2YBCO}
\end{figure}

To confirm that the ZBD observed in the $dI/dV$ spectrum at 4.2 K is due to the spin-polarization of CrO$_{2}$ and not to the spectral background of YBCO, we also measured YBCO/CrO$_{2}$ junction at 100K, above the $T_{c}$ of YBCO.  This normal-state data is plotted in the top left inset of Fig. \ref{CrO2YBCO}, and can be compared with the 4.2K data shown in the top right inset.  At 100K, YBCO is not superconducting and the $dI/dV$ spectrum, which does not show the kinks observed at 4.2 K, can be fitted over the entire voltage range using a polynomial.  At 4.2K, a similar polynomial can only fit the spectral regime for $e|V|$$>$$\Delta_0$.  For $e|V|$$<$$\Delta_0$, the measured $dI/dV$ deviates from the fit to the spectral background, indicating that the ZBD is in fact due to subgap spectral suppression.

It is worth noting that the YBCO/CrO$_{2}$ junction resistance at 4.2K ranged from 100 to 4000 $\Omega$. A high junction resistance $R$ indicates a small contact area, implying that the electron transport across the interface is highly local.  The effective point-contact radius $a$ can be calculated using the Wexler formula \cite{Wexler:927}, $R \approx 4\rho l/3\pi a^{2} + \rho /2a$, where $\rho $ is the residual resistivity and $l$ is the mean free path.  Using $\rho\approx$ 50$\mu \Omega$cm and $l\approx$ 10nm for YBCO \cite{Wei:PRL} and $R$ $\approx$ 0.1 - 4 k$\Omega$, our point contact radius was estimated to range from $a$ $\approx$ 0.7 to 6.0 nm, attesting to their ballistic ($a<l$) and microscopic nature.  Such small size of our YBCO point contacts suggests that they may be used to measure the spin polarization of individual magnetic domains.

To more clearly visualize the effect of spin polarization on YBCO point contacts, we compare the normalized $dI/dV$ spectrum taken on YBCO/CrO$_{2}$ from Figure \ref{CrO2YBCO} with a spectrum taken on YBCO/Au, as shown in Figure \ref{AuCrO2YBCO}.  For $e|V|$$>$$\Delta _0$ both spectra are relatively featureless.  For $e|V|$$<$$\Delta _0$ YBCO/Au shows a pronounced ZBP while YBCO/CrO$_{2}$ shows a distinct ZBD, with noticeable dips and kinks near $\Delta _0$.  To interpret these results more quantitatively, we performed spectral simulations using the spin-dependent $d$-wave BTK theory, as given by Refs. \onlinecite{spinpol_BTK} to \onlinecite{Zutic:PRB2}.  The left inset shows several $dI/dV$ spectra for various ``in-plane'' junction orientations and two values of $P$ at a fixed $Z$, illustrating the spectral variety for ideally oriented junctions.  The right inset shows the two corresponding angle-averaged spectra, each averaged within an Gaussian envelope centered on the $d$-wave node axis, to simulate our nominally-oriented (110) YBCO tip junctions.  There is good spectral resemblance between the YBCO/Au data and the $P$=0 simulation, and between the YBCO/CrO$_{2}$ data and the $P$=0.9 simulation.  These results confirm that the ZBP structure, which is formed by $d$-wave Andreev interference, is indeed suppressed by the high spin polarization of CrO$_{2}$ in our YBCO point-contact spectra.  

This work was supported by NSERC, CFI-OIT and the Canadian Institute for Advanced Research.  We thank Y.-T. Yen for technical assistance.


\begin{references}

\bibitem{BTK} G.E. Blonder, M. Tinkham, and T. M. Klapwijk, Phys. Rev. B \textbf{25}, 4515 (1982).
\bibitem{deJong:PRL} M.J.M. de Jong and C.W.J. Beenakker, Phys. Rev. Lett. \textbf{74}, 1657 (1995).
\bibitem{Soulen:science} R.J. Soulen Jr., J.M. Byers, M.S. Osofsky, B. Nadgorny, T. Ambrose, S.F. Cheng, P.R. Broussard, C.T. Tanaka, J. Nowak, J.S. Moodera, A. Barry, J.M.D. Coey, Science \textbf{85}, 282 (1998).

\bibitem{Upadhyay:3247} S.K. Upadhyay, A. Palanisami, R.N. Louie, and R. A. Buhrman, Phys. Rev. Lett. \textbf{81}, 3247 (1998).

\bibitem{Hu:94PRL1526} C.-R. Hu, Phys. Rev. Lett. \textbf{72}, 1526 (1994).
\bibitem{KT:95PRL351} Y. Tanaka and S. Kashiwaya, Phys. Rev. Lett. \textbf{74}, 3451 (1995).

\bibitem{spinpol_BTK} S. Kashiwaya, Y. Tanaka, N. Yoshida, M. R. Beasley, Phys. Rev. B \textbf{60}, 3572 (1999).
\bibitem{Zutic:PRB} I. Zutic and O. Valls, Phys. Rev. B \textbf{60}, 6320 (1999).
\bibitem{Zhu:PRB} Jian-Xin Zhu, B. Friedman, and C. S. Ting, Phys. Rev. B \textbf{59}, 9558 (1999).
\bibitem{Zutic:PRB2} I. Zutic and O. Valls, Phys. Rev. B \textbf{61}, 1555 (2000).

\bibitem{Vasko:844} V.A. Vas'ko, K.R. Nikolaev, V.A. Larkin, P.A. Kraus, and A.M. Goldman, Appl. Phys. Lett. \textbf{73}, 5774 (1998).
\bibitem{Chen:212508} Z.Y. Chen, Amlan Biswas, Igor Zutic, T. Wu, S.B. Ogale, R.L. Greene, and T. Venkatesan, Phys. Rev. B \textbf{63}, 212508 (2001).

\bibitem{Tsuei_revu} C. C. Tsuei and J.R. Kirtley, Rev. Mod. Phys.
\textbf{72}, 969 (2000).

\bibitem{CrO2_Parker} J. S. Parker, S. M. Watts, P. G. Ivanov, and P. Xiong, Phys. Rev. Lett. \textbf{88}, 196601 (2002).

\bibitem{Ranno:JAP} L. Ranno, A. Barry, and J.M.D. Coey, J. Appl. Phys. \textbf{81}, 5774 (1997).
\bibitem{Li:JAP} X.W. Li, A. Gupta, T.R. McGuire, P.R. Duncombe, and G. Xiao, J. Appl. Phys. \textbf{85}, 5585 (1999).
\bibitem{Gupta:APL} A. Gupta, X.W. Li, S. Guha, and G. Xiao, Appl. Phys. Lett. \textbf{75}, 2996 (1999).
\bibitem{Ivanov:JAP} P.G. Ivanov, S.M. Watts, and D.M. Lind, J. Appl. Phys. \textbf{89}, 1035 (2001).
\bibitem{Anguelouch:JAP} A. Anguelouch, A. Gupta, G. Xiao, G.X. Miao, D.W. Abraham, S. Ingvarsson, Y. Ji, C.L. Chien, J. Appl. Phys. \textbf{91}, 7140 (2002).

\bibitem{Woods:PRB} G. T. Woods, R. J. Soulen, Jr., I. Mazin, B. Nadgorny, M. S. Osofsky, J. Sanders, H. Srikanth, W. F. Egelhoff, and R. Datla, Phys. Rev. B \textbf{70}, 054416 (2004).

\bibitem{Strijkers:104510} G.J. Strijkers, Y. Ji, F.Y. Yang, C.L. Chien, J.M. Byers, Phys. Rev. B \textbf{63}, 104510 (2001).

\bibitem{footnote} We note that the BTK model used here largely neglects the Andreev evanescent wave, which could result in a small systematic correction to the calculated spectra \cite{Chalsani:PRB, Nadgorny:book}.

\bibitem{Ji:PRL} Y. Ji, G.J. Strijkers, F.Y. Yang, and C.L. Chien, Phys. Rev. Lett. \textbf{86}, 5585 (2001).


\bibitem{Geerk:329} J. Geerk, X.X. Xi, and G. Linker, Zeitschrift fur Physik B \textbf{73}, 329 (1988).
\bibitem{Sun:3266} A.G. Sun, L.M. Paulius, D.A. Gajewski, M.B. Maple, and R.C. Dynes, Phys. Rev. B \textbf{50}, 3266 (1994).

\bibitem{Wei:PRL} J.Y.T. Wei, N.-C. Yeh, D.F. Garrigus, and M. Strasik, Phys. Rev. Lett. \textbf{81}, 2542 (1998).

\bibitem{Alff:R14757} L. Alff, H. Takashima, S. Kashiwaya, N. Terada, H. Ihara, Y. Tanaka, M. Koyanagi, and K. Kajimura, Phys. Rev. B \textbf{55}, R14757 (1997).

\bibitem{Dagan:7012} Y. Dagan, A. Kohen, A. Revcolevschi, and G. Deutscher, Phys. Rev. B \textbf{61}, 7012 (2000).
\bibitem{Kashiwaya:RPP} S. Kashiwaya and Y. Tanaka, Rep. Prog. Phys. \textbf{63}, 1641 (2000).

\bibitem{Deu:Rewview} G. Deutscher, Rev. Mod. Phys. \textbf{77}, 109 (2005).


\bibitem{Hu:1266} C.-R. Hu, Phys. Rev. B \textbf{57}, 1266 (1998).
\bibitem{Wexler:927} G. Wexler, Proc. of the Phys. Soc. \textbf{89}, 927 (1966).


\bibitem{Chalsani:PRB} P. Chalsani, S. K. Upadhyay, O. Ozatay, and R. A. Buhrman, Phys. Rev. B \textbf{75}, 094417 (2007).

\bibitem{Nadgorny:book} B. Nadgorny, "Point Contact Andreev Reflection Spectroscopy", p. 531, in Handbook of Spin Transport and Magnetism, Taylor and Francis (2011), New York. 

\end{references}
\end{document}